\documentclass[fleqn,usenatbib]{mnras}

\usepackage{newtxtext,newtxmath}

\usepackage[T1]{fontenc}

\DeclareRobustCommand{\VAN}[3]{#2}
\let\VANthebibliography\thebibliography
\def\thebibliography{\DeclareRobustCommand{\VAN}[3]{##3}\VANthebibliography}

\usepackage{aas_macros}

\usepackage{graphicx}	
\usepackage{amsmath}	
\usepackage{mathrsfs}

\title[A model on radio NC]{An Empirical Estimation of the Galaxy Radio Number Counts model at 3~GHz and the Effect of Gravitational Lensing}

\author[K.T. Kono and T.T. Takeuchi]{
Kai T. Kono,$^{1}$\thanks{E-mail: kono.kai@c.mbox.nagoya-u.ac.jp (KTK)}
Tsutomu T. Takeuchi,$^{1,2}$
\\
$^{1}$Division of Particle and Astrophysical Science, Nagoya University,\\Furo-cho, Chikusa-ku, Nagoya 464--8602, Japan \\
$^{2}$The Research Center for Statistical Machine
Learning, the Institute of Statistical Mathematics, 10--3 Midori-cho, Tachikawa, Tokyo 190--8562, Japan
}

\date{Accepted XXX. Received YYY; in original form ZZZ}

\pubyear{2021}

\begin{document}
\label{firstpage}
\pagerange{\pageref{firstpage}--\pageref{lastpage}}
\maketitle

\begin{abstract}
We analyzed the evolution of 3GHz luminosity function of radio selected galaxies through cosmic time, up to about 1~Gyr after Big Bang, with making use of the VLA-COSMOS 3GHz large project. We estimated the luminosity function for SFGs and AGNs in the VLA-COSMOS field respectively with two independent methods, the non-parametric $C^-$ method for accuracy to avoid the effect of cosmic variance and parametric MCMC method assuming pure luminosity evolution. We obtained pure luminosity evolution parameter as $L_{*,{\rm SFG}}\propto (1+z)^{(3.36\pm0.05)-(0.31\pm0.02)z}$ for SFG and $L_{*,{\rm AGN}}\propto (1+z)^{(2.91^{+0.18}_{-0.19})-(0.99\pm0.08)z}$ for AGN. The resultant model agreed well with observed number counts down to sub-$\mu$Jy level. Additionally, we compared the model with cosmic star formation rate density history with multiple observations in various wavelength and confirmed that our result is consistent with literature up to $z\sim 3$. We further analyzed the statistical effect of gravitational lensing due to the dark matter halos and found that the effect is maximized at $S_{\rm 3\,GHz}\sim 100 \; {\rm \mu Jy}$ with about $1~$\% contribution for SFG and less than $0.5\%$ for AGN for the flux range considered in this study. In order to evaluate the effect of lensing magnification on parameter estimation in the SKA observations, we performed Fisher analysis and find that while the bias is limited, there is a distinguishable effect on SFG LF evolution parameter with SKA I MID wide survey, and all-sky and wide survey for AGN LF evolution parameters. 
\end{abstract}




\section{Introduction}
\label{sec:intro}

Exploring the galaxies evolution throughout cosmic time is the central topic of astrophysics. 
It dates back to 70's that astrophysicists started to pay attention to various evolutionary aspects of galaxies \citep[e.g.,][]{1974ApJ...192..293L,1978ApJ...219...46L}. 
Then, after the extensive studies by \cite{1980FCPh....5..287T} and coevals, the evolution has been regarded as one of the most fundamental properties of galaxies. 
Among a huge number of relevant studies, recent deep surveys in various wavelengths ~\citep[e.g.,][]{Smolcic2009, 2017A&A...602A...2S, Ueda2014, Padovani2015} and theoretical studies \citep[e.g.,][]{Croton2006,Hopkins2008,Somerville2008, 2015ApJ...810...72M, Caplar2015,Caplar2018} have shown a drastic evolution of physical properties of galaxies through the cosmic time. 

Especially, observations indicate that galaxies and super-massive black holes (SMBHs) at the center of galaxies have evolved together \citep[e.g.,][]{Marconi2003,2009ApJ...696..396S, 2009ApJ...696.1051K,Kormendy2013,2017ApJ...842...72Y,2018MNRAS.475.1887Y, 2018MNRAS.477.3030K}.  
This is referred to as the co-evolution of galaxies and SMBHs. 
These studies revealed that the star formation rate density (SFRD) have a peak at $z\sim 1\mbox{--}3$, sometimes called "the Cosmic Noon" \citep[e.g.,][among others]{Takeuchi2005,Bell2007, 2011A&A...528A..35M, Gruppioni2013,Burgarella2013,Madau2014, Bouwens2015, 2021ApJ...914..126M}. 
The accretion history onto SMBHs also has been constrained from observations of AGNs \citep[e.g.,][]{1982MNRAS.200..115S,2004MNRAS.351..169M,2007ApJ...654..754N,2018MNRAS.481.4971D}. 
For such approach, a luminosity function (LF) serves as a fundamental statistical tool to extract physical information from observations \citep[e.g.,][]{2004MNRAS.351..169M,2007ApJ...669..776K,2008ApJ...679..118S,2016ApJ...820...65Y,2017ApJ...842...72Y}. 

The radio LF has been studied for many years based on the data mainly from the VLA \citep[e.g.,][]{Machalski2000, Mauch2006, Smolcic2009}. 
These studies revealed that the local radio LFs consist of two distinct  components. 
The faint end of LF is dominated by star forming galaxies (SFG) and the bright end is dominated by active galactic nuclei (AGN). 
The radio continuum emission from SFG is due to synchrotron emission from relativistic electron accelerated by the shock of supernova remnants. 
Therefore, the radio luminosity from SFGs can be interpreted as the tracer of star forming rate (SFR) of the galaxy \citep{Condon2008, Murphy2011}. 
On the other hand, the radio emission from AGNs is dominated by the synchrotron emission from Mpc scale radio lobe and associated jet from center of a galaxy. 
Hence, the radio emission from AGNs can be interpreted as the mass accretion rate onto a SMBH at the center of galaxies \citep[e.g.,][]{1982MNRAS.200..115S,2004MNRAS.351..169M,2007ApJ...654..754N,2008MNRAS.383..277K,2018MNRAS.481.4971D}. 
AGNs play an important role to understand the evolution of galaxies through radio mode AGN feedback \citep[e.g.,][]{Croton2006,2007ApJ...669..776K, 2008MNRAS.388.1011M,Fabian2012}.

Since the estimation of a LF requires redshift measurements, it is not immediately available from the completion of a certain survey. 
Observed number counts of galaxy provides the simplest statistics to constraint their evolution. 
It had been used to constraint cosmological parameters in earlier studies as the value is sensitive to the geometry considered.  
\cite{2000ApJS..129....1T, Takeuchi2001} investigated FIR source counts and pointed out that strong luminosity evolution is necessary to explain the {\it IRAS} source counts. 
In radio band, the situation is slightly different as the bright end of radio number count is dominated by AGNs, expecially referred as radio loud AGNs, while most of the faint end population comes from SFGs \citep[e.g.,][]{1984ApJ...287..461C, 2011A&A...536A..13P, Padovani2015, 2016MNRAS.462.2934V}. \citep[see also][]{2016A&ARv..24...13P}.
Additionally, probability of deflection, or $P(D)$ analysis is also applied to achieve confusion-limited number counts of galaxies below survey sensitivity limit which is pioneered by \cite{1957PCPS...53..764S}. The $P(D)$ is the probability distribution of peak flux and consists of the convolution of system noise, residual sidelobes and so on. 
This analysis is widely applied in various observation frequency such as in infrared band \citep[e.g.,][]{2004ApJ...604...40T,2011A&A...532A..49B}  and  in radio and submillimeter band \cite[e.g.,][]{2012ApJ...758...23C,2014MNRAS.440.2791V, 2020ApJ...897...91G, 2020ApJ...888...61M}. 
Especially, the radio number counts is investigated down to sub- $\mu{\rm Jy}$ in \cite{2014MNRAS.440.2791V} and \cite{2020ApJ...888...61M} with $P(D)$ analysis.

For the analysis of galaxy/AGN number counts, we have to take into account yet another important effect, gravitational lensing. 
It is well known that the galaxy number counts in submillimeter band shows prominent magnification effect of gravitational lensing. 
Some previous studies show that approximately $60~\%$ of galaxies are strongly lensed in the bright end \citep[e.g.,][]{Lima2009,Bethermin2012,Negrello2017,2017ApJ...842...95M}  
in the submillimeter and millimeter band as a result of the negative $K$-correction because of the shape of their spectral energy distribution. 
This effect causes a significant bias when we estimate the evolution of galaxy from blind survey data. 
Furthermore, the development of very long base line interferometers (VLBI) such as square kilometre array (SKA) and the Karl G. Jansky Very Large Array (VLA) allow us to perform deep survey even in the radio band. 
Distant galaxies are more likely to be lensed due to the long path toward us. 
We therefore need to estimate the statistical effect of gravitational lensing on the galaxy number counts. 


In this study, we estimated the radio LFs for $0.1<z<5.5$ at 3GHz based on the data obtained from the VLA-COSMOS {\rm 3GHz} large project \cite{2017A&A...602A...1S}, and construct a galaxy evolution model that is able to reproduce observed number counts with assuming  the pure luminosity evolution (PLE) model that only the parameters on luminosity vary with redshift. Afterwards, we discuss about properties future SKA and other panchromatic surveys based on resultant LF evolution model. \par

The structure of this paper is as follows. 
In Section~\ref{sec:data}, we outline the VLA-COSMOS data and the classification of the sample. 
In Section~\ref{sec:LF}, we show the method to estimate the radio LFs. In section~\ref{sec:res}, we show the results of the estimated LFs and construct our evolution model From Section~ \ref{sec:SFRD} to \ref{sec:GL}, we discuss about the effect of gravitational lensing on galaxy number counts and cosmic star formation rate density history based on obtain LF evolution for SFG. 
Finally, we summarize our results in section~\ref{sec:conclusion}. \par

In this study, we adopt cosmological parameter with $H_0=70\ [{\rm km\ s^{-1}\ Mpc^{-1}}]$, $\Omega_{\Lambda}=0.7$, $\Omega_{M}=0.3$, $\sigma_8=0.8$, throughout this paper.

\section{DATA}
\label{sec:data}
\subsection{Radio sample}
The Cosmic Evolution Survey (COSMOS) field \citep{Scoville2007} covers about $2\ {\rm deg}^2$ area on the sky whose center is $({\rm R.A.},{\rm decl.})=(150.1163213, 2.20973097)$ with panchromatic deep observations data from radio to X-ray. 
We analyzed the data taken from the VLA-COSMOS 3GHz large project \citep{2017A&A...602A...1S}.  
The rms sensitivity is $2.3\ {\rm \mu Jy}\ {\rm beam}^{-1}$ with angular resolution of $0^{\prime \prime}. 75$ and the observation time is 384 hours in the VLA S-band. The whole area of the VLA-COSMOS observation is $2.6 {\rm deg}^2$. 
They masked the area with bright NIR/optical counterparts and achieve the effective area with $1.77\ {\rm deg^2}$. 
The 10,830 objects with ${\rm S/N}\geq 5$ has been identified within the effective area. 
Since the rms sensitivity is $2.3\ {\rm \mu Jy}\ {\rm beam}^{-1}$, $5\sigma$ detection corresponds to flux limit $S_{\rm lim}=11.0\ {\rm \mu Jy}$ in this study. 
These objects are cross matched with near-infrared (NIR)/optical sources in COSMOS2015. 
COSMOS2015 catalog contains over 800,000 sources observed in bands from near ultra violet (NUV) to radio band taken by GALEX, UltraVISTA DR2, Subaru/HyperSuprime-Cam, and the SPLASH Spitzer legacy program in main catalog, and supplementary provide survey data in X-ray taken by {\it Chandra}, {\it NuStar} and {\it XMM-Newton} and in radio band by VLA  (see \cite{Laigle2016} in detail).

The photometric redshift is available in 7826 sources in VLA-COSMOS 3GHz data and spectroscopic redshift is available for 2740 objects. We made use of the best redshift available either photometric or spectroscopic redsfhift. 

The sample classification has done in \cite{2017A&A...602A...1S}  and \cite{Delvecchio2017}. They investigate the multiwavelength data of AGN host galaxies in the COSMOS field and find that galaxies with $3\sigma$ of radio excess are classified by the criteria below, with $\rm SFR$ estimated from integrated infrared (IR) luminosity ${\rm SFR}_{\rm IR}$ and $1.4{\rm GHz}$ luminosity $L_{\rm 1.4 GHz}$,
\begin{equation}
\label{eq:Rex}
    \log \left(\frac{L_{1.4 \mathrm{GHz}}\left[\mathrm{W}\  \mathrm{Hz}^{-1}\right]}{\mathrm{SFR}_{\mathrm{IR}}\left[\mathrm{M}_{\odot}\  \mathrm{yr}^{-1}\right]}\right)>a(1+z)^{b}
\end{equation}
where $a=22.0$ and $b=0.013$. The IR luminosity is estimated in \cite{2017A&A...602A...2S} and they assume \cite{KennicuttJr.1998} relation and Chabrier IMF \citep{2003PASP..115..763C} to convert $L_{\rm IR}$ to ${\rm SFR}_{\rm IR}$. These objects refer to galaxies more than $80\%$ of whose luminosity is owed by AGN. Color selection has done in rest frame color and they defined clean SFG as SFGs whose radio excess is below the criteria in Eq.~\ref{eq:Rex}. Under this criteria, more than $80\%$ of the radio emission is owed by AGN radiation \citep{2017A&A...602A...6S}.
Finally, the selected number of SFG is 5410 and the number of AGN is 1908.




\section{Method : Luminosity function estimation}
\label{sec:LF}
Some literature have already investigated the evolution of radio luminosity function in the COSMOS field \citep[e.g.,][]{Smolcic2009, 2017A&A...602A...6S, 2017A&A...602A...5N,Novak2018, 2018A&A...620A.192C, 2021ApJ...907....5V, 2021MNRAS.tmp.2958M} on several types of SFG and AGN up to $z\lesssim 6$. 
In this paper, we provide evolution of luminosity function up to $z\lesssim 5.5$ estimated from two independent methods. 
One is a non-parametric estimator called the $C^-$ method \citep{1971MNRAS.155...95L} and the other one is parametric Bayesian estimator which is formulated in \citep{2008ApJ...682..874K}.
\subsection{$K$-correction}
Expansion of the Universe causes redshift in observed SED with respect to the rest-frame SED. 
Especially in high redshift survey, we need to take the effect into account which is called $K\mbox{-}$ correction. 
The monochromatic luminosity $L_{\nu}$ as a function of frequency $\nu$ for a galaxy at redshift $z$ can be written as below,
\begin{equation}
    \label{eq:K_corr}
    L_{\nu_{\rm em}}=L_{\nu_{\rm obs}(1+z)}=\frac{4\pi d_L(z)^2S_{\rm obs}}{(1+z)}
\end{equation}
where $\nu_{\rm obs}$ is the observed frequency, $S_{\rm obs}$ is the observed flux at $\nu_{\rm obs}$ and $\nu_{\rm em}$ is the rest flame frequency at the source. 
We assumed the single power law for radio spectra as $S_{\nu}\propto \nu^{-\beta}\ (\beta>0)$, where $\beta$ is the spectral index since the low frequency radio spectra is mainly dominated by synchrotron radiation from relativistic electron which is accelerated at shock front of super novae remnants for SFGs and at jet and lobe for AGNs. 
Because of this, the $K$-correction for radio band can be descried by simple power law. We made use of the flux at 1.4GHz from the VLA-COSMOS 1.4GHz survey \citep{2006ApJ...649..181S} to derive the spectral index of each galaxy. 
For sources that have counterparts in 1.4GHz, we employed the spectral index determined by the data. 
The average values for each population are $\langle \beta_{\rm SFG} \rangle=0.87\pm 0.15$, $\langle \beta_{\rm AGN} \rangle=0.85 \pm 0.26$. 
The flux limit for this region is $S_{\rm lim}=11.0\ [{\rm \mu Jy}]$ which is correspond to the $5\sigma$ detection. For sources without 1.4GHz counterparts, we assumed the average values $\langle \beta_{\rm SFG}\rangle$ and $\langle \beta_{\rm AGN}\rangle$ as spectral index for each galaxy type.

\subsection{$C^-$ method}
We adopt the $C^-$ method \citep{1971MNRAS.155...95L} to estimate the radio LFs in this study. 
It is a non-parametric method and is mathematically verified that the method is able to avoid the effect of cosmic variance \citep[e.g.,][]{1997AJ....114..898W, 2000ApJS..129....1T}. 
Therefore, the $C^-$ method is a suitable method to estimate accurate LFs for a deep survey which can be easily biased by cosmic variance because of its small survey area. 
The $1/V_{\rm max}$ method \citep{1968ApJ...151..393S} is the most commonly adopted method to obtain non-parametric luminosity function. 
Although, since $1/V_{\rm max}$ method  assumes isotropic special distribution in the input samples, the LF obtained by $1/V_{\rm max}$ method can be influenced by the presence of clusters or voids \citep{1997AJ....114..898W, 2000ApJS..129....1T}. 
The $C^-$ method gives the cumulative luminosity function $\Phi$, calculated as below.
\begin{equation}
    \Phi(L) \propto \sum^{L_k<L}_{k=1}\psi_k=\psi_1 \prod^{L_k<L}_{k=1} \frac{C^-_k+1}{C^-_k}
\end{equation}
The value $C^-_k$ is defined as the number of samples in a region $(L_k, L_ u]\times [z_l,z_{u(k)}]$, where $L_k$ and $z_k$ denotes the luminosity and the redshift of $k$ th galaxy, respectively, and $z_{{\rm max}(k)}$ denotes the maximum redshift which the $k$ th galaxy can be observed with a certain flux limit $S_{\rm lim}$, $L_u$ is the upper limit of the luminosity range and $z_l$ is the lower limit of the redshift range. 
Note that the luminosity is sorted in ascending order.
With the cumulative LF $\Phi(L)$, differential LF $\phi(L)$ can be determined as follows for logarithmic intervals.
\begin{equation}
    \phi(L)=\frac{d\Phi}{d\log L}
\end{equation}
In this study, we divided the whole sample into 11 redshift bins from $z=0.1$ to $z=5.7$, the same manner in \cite{Novak2018} and applied $C^-$ method for each of them.


\subsection{Estimation of Luminosity Function with MCMC}
\begin{figure*}
\begin{center}
\includegraphics[width=0.98\textwidth]{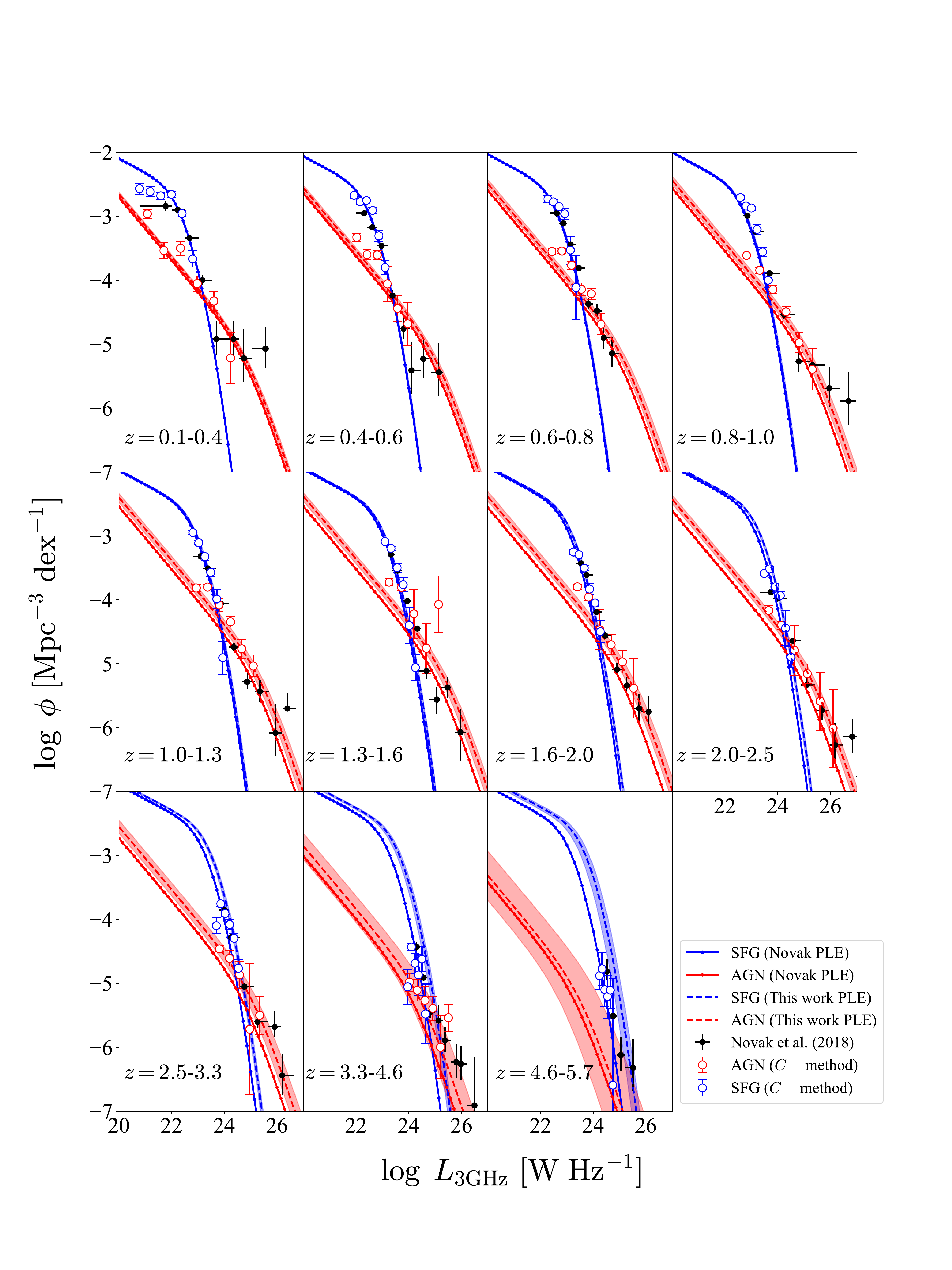}
\caption{The radio LF of star forming galaxies and AGNs for nine redshift ranges evaluated with $C^-$ method. Blue points and lines shows star forming galaxies and red points and lines show AGNs respectively. The shaded area are the one sigma error of estimated parameters.}
 \label{fig:LF}
\end{center}
\end{figure*}

\begin{table*}
    \centering
    \begin{tabular}{c||cccc} 
         & $\alpha^{\rm SFG}_1$ & $\alpha^{\rm SFG}_2$ & $\alpha^{\rm AGN}_1$ & $\alpha^{\rm AGN}_2$ \\ \hline
    This work                        & $3.36\pm 0.05$ & $-0.31\pm 0.02$ & $2.91^{+0.18}_{-0.19}$ & $-0.99\pm 0.08$ \\ \hline
     \cite{Novak2018} (3~GHz)& $3.16\pm 0.04$ & $-0.32\pm 0.02$ & $2.88\pm 0.17$ & $-0.84\pm 0.07$  \\ \hline
    \cite{McAlpine2013} (1.4~GHz)     & $2.47\pm 0.12$ &                 & $1.18\pm 0.21$ &  \\ \hline
    \end{tabular}
    \caption{Comparison of best fit PLE parameters.}
    \label{tab:res_paras}
\end{table*}

The Markov chain Monte Carlo (MCMC) method is a sampling algorithm from multivariate probability distribution. 
We made use of the Python package \texttt{EMCEE} \citep{2013PASP..125..306F} to perform MCMC sampling and multi-variate parameter fitting to our data in this study.
Adopting the formulation introduced \cite{2008ApJ...682..874K, 2013MNRAS.428..291P}, the probability of finding $i\mbox{-}$th galaxy at $\{L_i, z_i\}$ with the luminosity range $[\log{L},\log{L}+\mathrm{d} \log{L}]$ and redshift range and $[z,z+\mathrm{d} z]$ for given parameter set $\{\theta\}$ is,
\begin{equation}
    p(L, z \mid\{\theta\})=\frac{\phi(L, z \mid\{\theta\}) p({\rm selected } \mid L, z)}{\lambda} \frac{\mathrm{d} V}{\mathrm{~d} z}
\end{equation}
where $\lambda$ is the expected number of galaxies which is calculated from evolution model of luminosity function, $\phi(L,z)$ is luminosity function at redshift $z$,  survey configuration and $dV/dz$ is volume element. 
\begin{equation}
\begin{split}
    \lambda=&\sum_{\rm fields} \iint \phi(L, z \mid\{\theta\}) p({\rm selected}  \mid L, z) \mathrm{d} \log L \frac{\mathrm{d} V}{\mathrm{d} z} \mathrm{d} z\\
    &=4\pi \frac{\mathcal{A}_{\rm sky}}{\mathcal{A}_{\rm all}} \int^{\infty}_{L_{\rm lim}}\int^{z_{\rm max}}_{z_{\rm min}} \phi(L, z \mid\{\theta\}) C(S(L))C(z) \mathrm{d} \log L \frac{\mathrm{d} V}{\mathrm{d} z} \mathrm{d} z
\end{split}
\end{equation}
where $C(S)$ is the completeness and bias correlation factor for the VLA-COSMOS 3GHz catalog, $\mathcal{A}_{\rm sky}$ is effective sky area and $\mathcal{A}_{\rm all}$ is all sky area respectively.
Effective sky area is $1.77\ {[\deg]}$ in VLA-COSMOS 3GHz Large Project. 
Here, $p({\rm selected}  \mid L, z)$ is selection function defined as,
\begin{equation}
    p({\rm selected}  \mid L, z) = p({\rm det}  \mid L, z)p(L>L_{\rm lim}\mid L,z)
\end{equation}
where $p({\rm det}  \mid L, z)=C(S)C(z)$ which indicates the detection probability of the observation and $p(L>L_{\rm lim}\mid L,z)$ that restricts integration range for luminosity. In this study, we adopted the value in Table 2 in \cite{2017A&A...602A...1S} for $C(S)$. 
As for $C(z)$, we calculated the rate that a galaxy has its counterpart in other bands at given redshift with referring to optical and NIR survey in COSMOS2015 \cite{2016ApJS..224...24L}.
We assumed that the number of detected sources follows Poisson distribution with expected galaxy number counts $\lambda$. With these assumptions, the likelihood function is,
\begin{align}
&p\left(N,\left\{L_{i}, z_{i}\right\} \mid\{\theta\}\right) \propto \nonumber \\
&\quad\lambda^{\sum w_{i}} \mathrm{e}^{-\lambda}
\prod_{i=1}^{N}\left\{\frac{\phi(L_i, z_i \mid\{\theta\}) p({\rm selected } \mid L_i, z_i)}{\lambda} \frac{\mathrm{d} V}{\mathrm{~d} z}\right\}^{\frac{w_i}{\langle w \rangle}}
\end{align}
where $w_i$ is the incompleteness of the $i\mbox{-}$th galaxy and $\langle w\rangle$ is the average of incompleteness.
Generally, the cosmological evolution of the luminosity function is described as a combination of luminosity evolution and density evolution. 
\begin{equation}
    \phi(L,z)=g(z)\phi_0{\left[\frac{L}{f(z)}\middle| \{\theta\}\right]}
\end{equation}
where $f(z)$ is luminosity evolution, $g(z)$ is density evolution term respectively and $\phi_0$ indicates luminosity function at $z=0$. 
Assuming only $f(z)$ or $g(z)$ is referred as pure luminosity evolution (PLE) and pure density evolution (PDE) which is often employed in the literature \citep[e.g.,][]{1984ApJ...284...44C, Takeuchi2001, Gruppioni2013}. 
This time, we assumed the evolution function as whose power is first order polynomial for redshift. 
Some study proposed luminosity dependent density evolution model \citep[PDDE; ][]{2003ApJ...598..886U,Ueda2014} which is motivated from studies on SMBH downsizing. 
However, smallness of number of our sample did not allowed us to obtain sufficient constraint on PDDE parameters. 
On the other hand, as pointed out in \cite{Novak2018}, PDE model would overestimate LF especially at low luminosity and we obtained the similar result when considering the mixture of parametric PLE and PDE model.
Based on this, we assumed PLE model with two parameters $\alpha_0,~\alpha_1$ in this study.
\begin{equation}
    f(z)=(1+z)^{Q(z)}, \quad \quad Q(z) = \alpha_0+\alpha_1 z
\end{equation}

We assumed Saunders function \cite{1990MNRAS.242..318S}, a combination of power-law and lognormal distribution with four parameters ($\phi_0$, $L_*$, $\alpha_{\rm S}$, $\sigma$) for SFG luminosity function which is often applied.
\begin{equation}
    \phi(L)=\phi_*{\left(\frac{L}{L_*}\right)}^{1-\alpha}\exp{{\left[-\frac{1}{2\sigma^2}\log^2{\left(1+\frac{L}{L_*}\right)}\right]}}
\end{equation}

As for AGN luminosity function, we assumed double power law function with four parameters ($\phi_0$, $L_*$, $p_1$, $p_2$).
\begin{equation}
    \phi(L)=\frac{\phi_*}{(L/L_*)^{p_1}+(L/L_*)^{p_2}}
\end{equation}
We adopt the parameters at $z=0$ which is obtained by \cite{2007MNRAS.375..931M}. The value of parameters are, $\phi_{*,\ \rm SFG}=10^{-2.45\pm 0.05}\ [{\rm Mpc^{-3}\ dex^{-1}}]$, $L_{*,\ \rm SFG}=10^{21.26\pm 0.22}\ [{\rm W\ Hz^{-1}}]$, $\alpha=1.02\pm 0.15$, $\sigma=0.60\pm 0.04$ for SFG LF and $\phi_{*,\ \rm AGN}=\frac{1}{0.4}10^{-5.5\pm 0.25}\ [{\rm Mpc^{-3}\ dex^{-1}}]$, $L_{*,\ \rm AGN}=10^{24.59\pm 0.30}\ [{\rm W\ Hz^{-1}}]$, $p_1=1.27\pm 0.18$, $p_2=0.49\pm 0.04$ for AGN LF. All the parameters on LF slopes ($\alpha$ and $\sigma$ for SFG, and $p_1$ and $_2$ for AGN) are fixed in parametric estimation as the sample luminosity is limited to brighter than $L_*$ for most of redshift bins.  In this MCMC calculation, we assumed that prior probability function $p(\theta)$ as step function in range a of $-20\leq \alpha_{L,1,{\rm SFG}} \leq 20$, $-20\leq \alpha_{L,2,{\rm SFG}} \leq 20$, $-20\leq \alpha_{L,1,{\rm AGN}} \leq 20$ and $-20\leq \alpha_{L,2,{\rm AGN}} \leq 20$.

\section{Results}
\label{sec:res}
\subsection{The estimated radio luminosity functions}
\begin{figure*}
\begin{center}
\includegraphics[width=7.5cm]{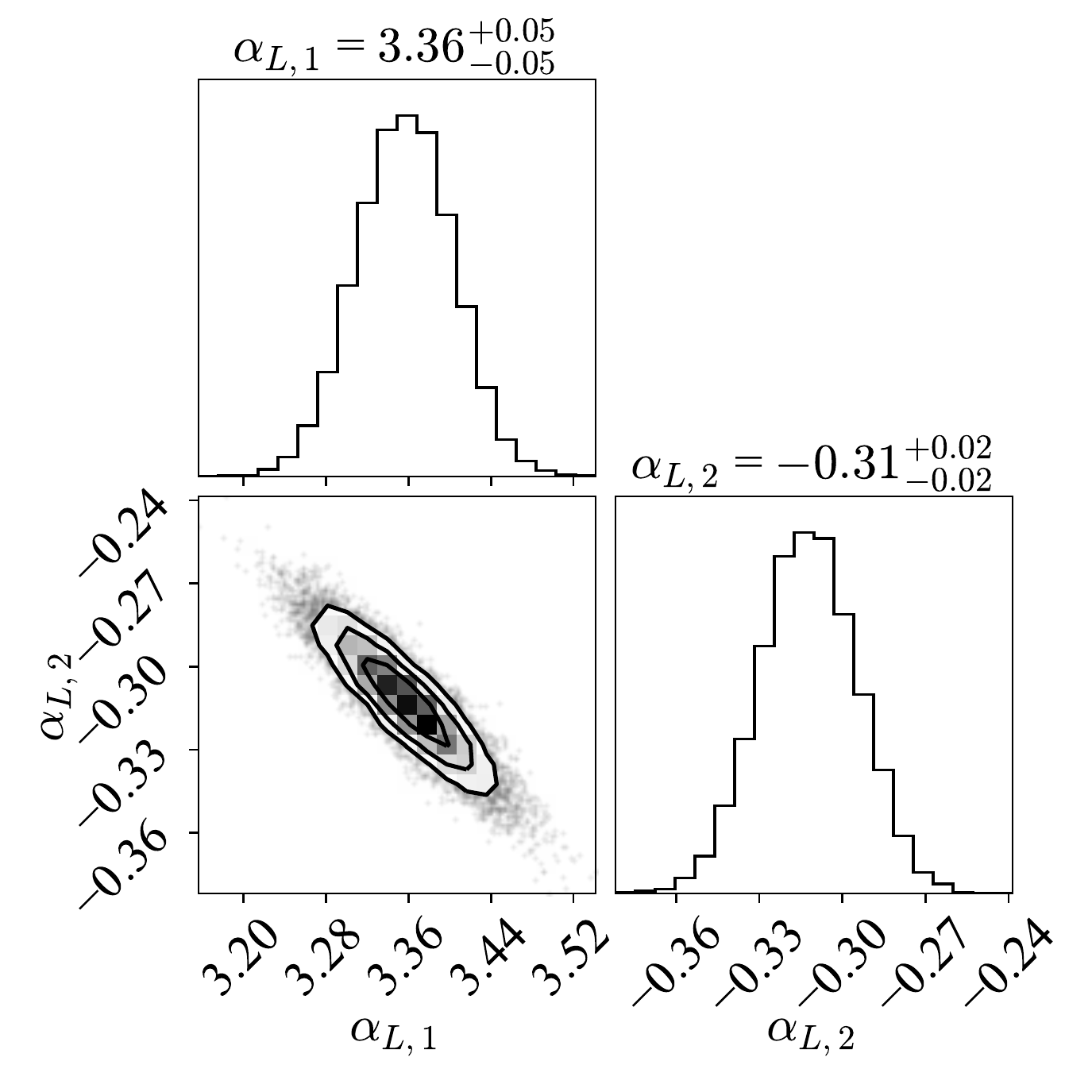}
\includegraphics[width=7.5cm]{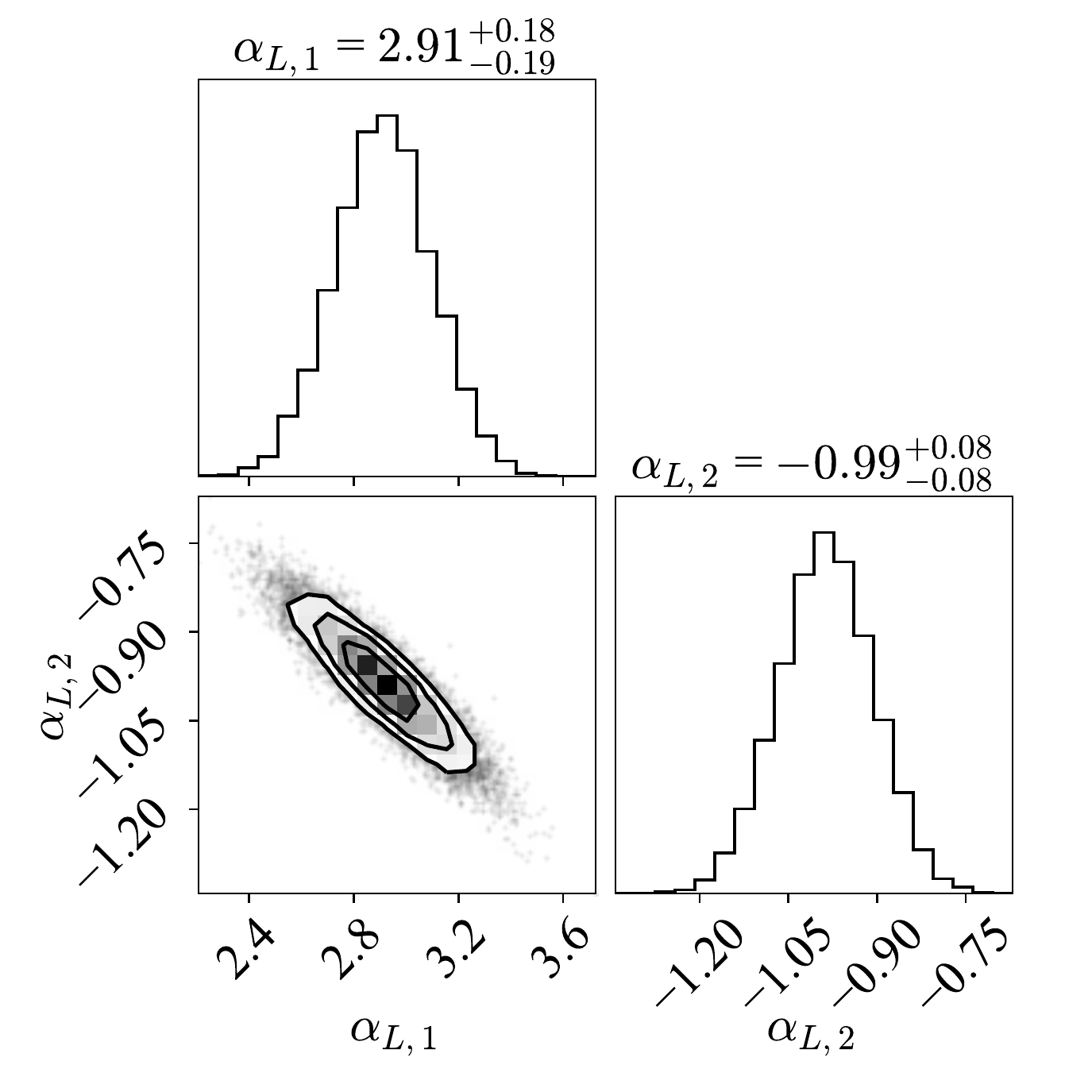}
\caption{Constrains of evolution parameter $\alpha_{L,1}$ and $\alpha_{L,2}$ with MCMC for SFG (left) and for AGN (right).}
\label{fig:MCMC}
\end{center}
\end{figure*}

The obtained LFs for all 11 redshift bins are displayed in Fig.~\ref{fig:LF} along with parametric and non-parametric LFs obtained in \cite{Novak2018}. 
The error bars for points obtained through $C^-$ method is estimated with boot strapping re-sampling method with $95\%$ confidence region.
The parameters for the first redshift range are not well constrained since the number of the sample is not enough because of the volume effect. 
In Tab.~\ref{tab:res_paras}, we display our resultant best fit parameters along with LF parameters obtained in previous works for comparison. Our results are consistent with those obtained in \cite{Novak2018} in almost all parameters. \cite{McAlpine2013} estimates the PLE parameters from $1.4~{\rm GHz}$ radio LFs with the sources in the VIDEO-XMM3 field of SFG and AGN up to $z \sim 2.5$. While they assumed constant $Q(z)$, the values averaged over corresponding redshift range are consistent with our result. As shown in Fig.~\ref{fig:MCMC}, $\alpha_1$ and $\alpha_2$ have negative correlation for both of SFGs and AGNs.

\subsection{Number counts}
\label{Sec:nc}
\begin{figure*}
\begin{center}
\includegraphics[width=\textwidth]{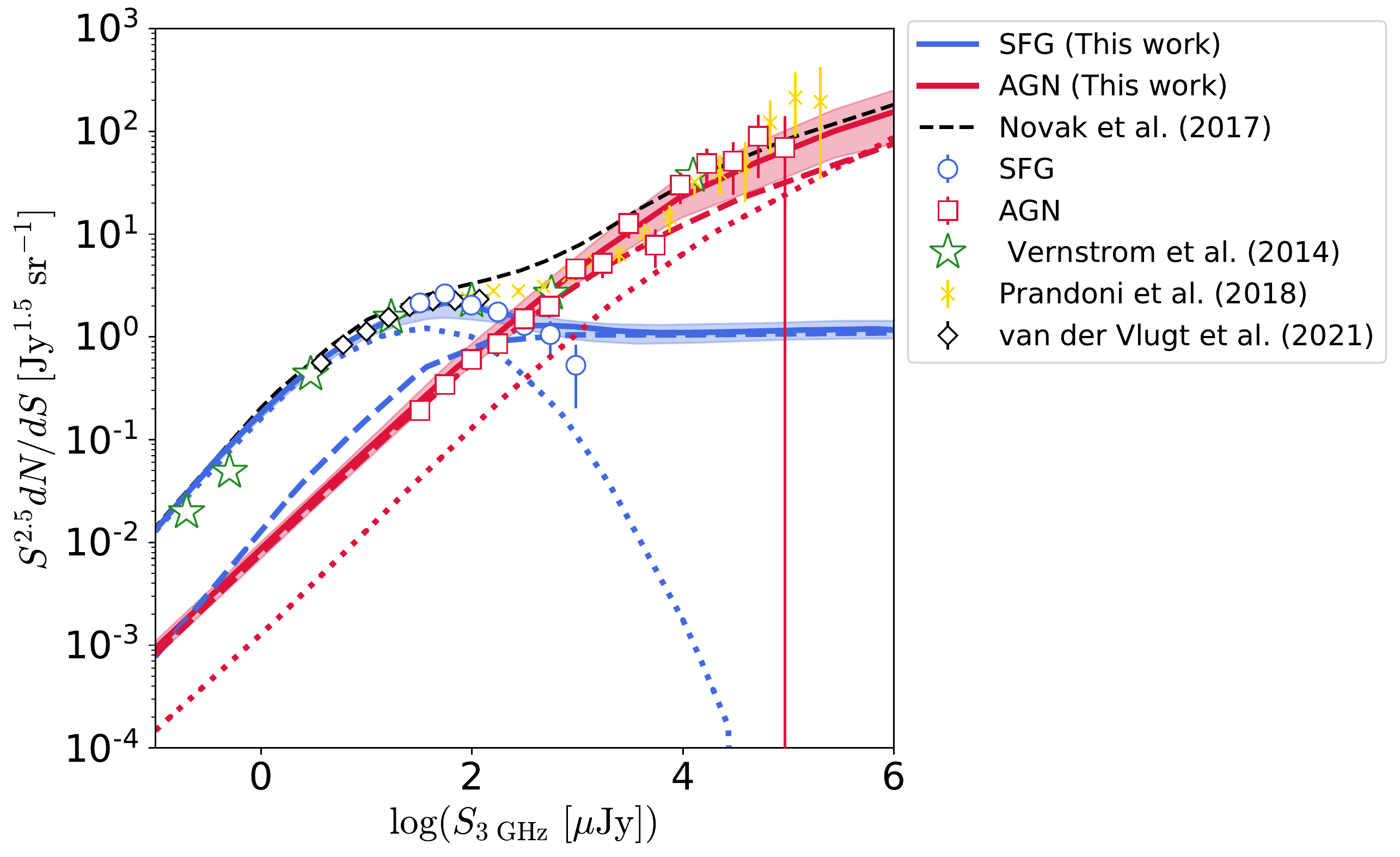}
\caption{The Euclid normalized differential number counts at $3~{\rm GHz}$. Blue circles indicate observed SFG number counts and red squares indicate AGN, respectively. The error bars corresponds to $1\sigma$ error considering the effect of galaxy clustering since the survey are is not sufficiently large \citep{Takeuchi2001}. We assumed the correlation function as $\omega (\theta)\propto \theta^{-0.8}$ obtained by \citet{Hale2018}. Blue and red solid lines and shaded area indicate the differential number counts obtained by the PLE model of SFG and AGN with $95\%$ confidence band. Dashed and dotted lines indicates contribution from low/high redshift population separated at $z=0.5$ for each galaxy species. Black dashed line indicates PLE model in \citet{Novak2018}. Green stars:\citet{2014MNRAS.440.2791V} yellow asterisks:\citet{2018MNRAS.481.4548P}, black diamonds: \citet{2021ApJ...907....5V} are observed points in radio survey.}
\label{fig:diff}
\end{center}
\end{figure*}

With the redshift dependent LF, the radio galaxy number counts $N(>S)$ can be calculated.
The galaxy number counts per unit solid angle with redshift dependent luminosity function, $\phi(L,z)$, detected greater than $S_{\rm lim}$ can be formulated as below,
\begin{equation}
    N(>S)= \int^{z_{\rm max}}_0dz\frac{dV}{dz}\int^{\infty}_{L(z,S_{\rm lim})}\phi(L,z)dL
\end{equation}
The  lower limit luminosity $L(z,S_{\rm lim})$ in the second integration is calculated from Eq.~(\ref{eq:K_corr}). 
The Euclid normalized differential number counts calculated from our model are displayed in Fig.~\ref{fig:diff} along with results from other radio surveys.
The errors for the differential number counts are larger than that is shown in previous studies \cite{2017A&A...602A...2S}. As pointed out by \cite{Takeuchi2001}, Poisson statistic underestimates the errors on galaxy number counts as the effect of galaxy clustering on variance of number count is not negligible. The total variance of galaxy number counts $N$ is written as, 
\begin{equation}
    \langle (N-\langle N \rangle)^2 \rangle \simeq \mathcal{N}\Omega+\mathcal{N}^2\Omega \int_{\Omega} \omega (\theta) d\Omega
\end{equation}
where $\mathcal{N}$ is the surface number density, $\Omega$ is the observed area and $\omega (\theta)$ is the angular correlation function. In this study, we made use of angular correlation functions for VLA-COSMOS sample obtained by \cite{Hale2018}. We adopted correlation function as, $\omega(\theta)=10^{-2.88}\theta^{-0.8}$ for SFG and $\omega(\theta)=10^{-2.57}\theta^{-0.8}$ for AGN, respectively.

Our model successfully reproduce the observed number counts in all flux range comparing with result from \cite{2014MNRAS.440.2791V}, \cite{2018MNRAS.481.4548P}, \cite{Novak2018} and  \cite{2021ApJ...907....5V}. 
\cite{2018MNRAS.481.4548P} number counts is obtained by $1.4~{\rm GHz}$ observation with Westerbork Synthesis Radio Telescope in the Lockman Hole.
The result from \cite{2014MNRAS.440.2791V} obtained by making use of $P(D)$ analysis based on VLA $3~{\rm GHz}$ observation in the Lockman Hole. They derived deifferential number counts down to $50~{\rm nJy}$.
In \cite{2021ApJ...907....5V}, they performed ultra-deep observation, the COSMOS-XS survey with sensitivity of $0.53~{\rm \mu Jy~beam^{-1}}$ in band S ($3~{\rm GHz}$) and $0.41~{\rm \mu Jy~beam^{-1}}$ in band X ($10~{\rm GHz}$). 
Their effective survey area is $180~{\rm arcmin}^2$ and $16~{\rm arcmin}^2$, respectively.
Our result succeed to reproduce their sub-${\rm \mu Jy}$ number counts as well.
The number counts below $S_{\rm 3GHz}=0.1\ {\rm mJy}$ is dominated by SFGs while the bright end is dominated by AGNs which is consistent with conventional understanding. 

We extrapolated our result and evaluated differential number counts down to $100~[{\rm nJy}]$ and showed that about $95\%$ of faint radio population is dominated by SFG.
We also display the contribution from low and high redshift separated at $z=0.5$. 
The plateau in bright end of SFG number counts is mainly contributed by nearby galaxies at $z<0.5$ which is indicated by blue dashed line in Fig.~\ref{fig:diff}. Blue dotted lines indicates that $90\%$ of faint end SFG population comes from high-z.
In contrast to this, approximately $50\%$ of bright end of AGN is dominated by galaxies at $z>0.5$.
The population fraction as a function of flux is fully consistent with the Fig.5 in \cite{Novak2018}. \par


\begin{figure*}
\begin{center}
\includegraphics[width=\textwidth]{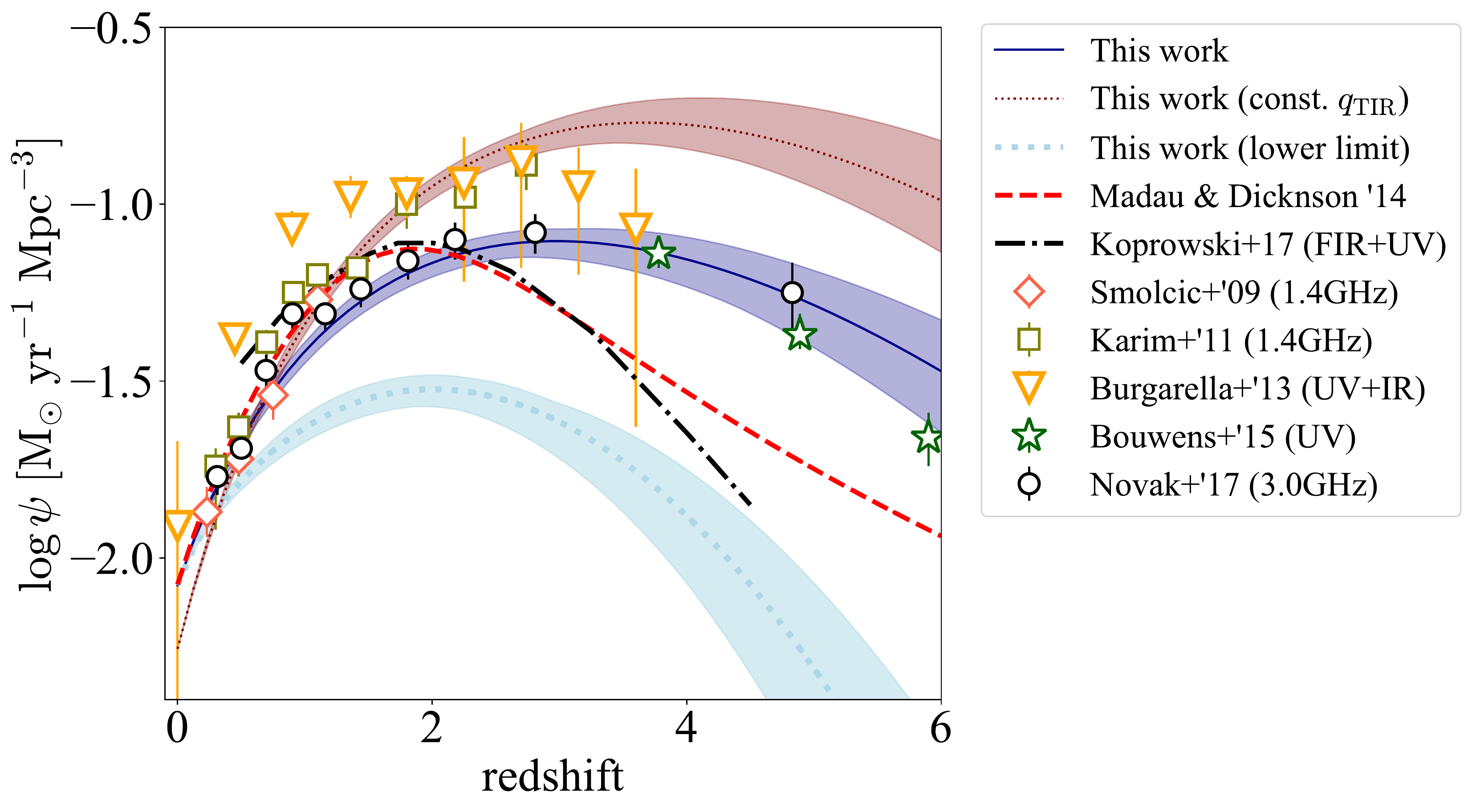}
\caption{Redshift evolution of cosmic SFRD up to $z\sim 6$. The blue line and shaded region show the result of this work with $2\sigma$ error. 
The light blue dotted line along with shaded area indicates our lower limit whose integral base assumed to be $L_{\rm lim}(z,S_{\rm lim})$ that is given in Eq.~\ref{eq:K_corr}.
The brown dotted line is obtained by assuming constant $q_{\rm TIR}$.
The red diamonds denotes SFRD obtained from observation in ${\rm 1.4~GHz}$ \citet{Smolcic2009}, yellow squares are from  ${\rm 1.4~GHz}$ \citet{Karim2011}, orange triangles are from combination of IR and UV \citet{Burgarella2013}, green stars are from dust corrected UV \citet{Bouwens2015} and black circles are from  ${\rm 3~GHz}$ \citet{2017A&A...602A...5N}. The red dashed line is the analytical form obtained by \citet{Madau2014} and the black dash-dotted line is obtained from FIR observation \citet{2017MNRAS.471.4155K}.}
\label{fig:sfrd}
\end{center}
\end{figure*}

\section{Cosmic SFRD}
\label{sec:SFRD}

We calculated the SFRD up to $z\sim 6$ based on the radio continuum luminosity density for SFGs with our evolution model. We made use of a tight correlation between radio and IR emission \citep[e.g.,][]{1992ARA&A..30..575C, Yun2001} to convert $L_{\rm 3GHz}$ to SFRD. The relation commonly parameterized with $q\mbox{-}$parameter introduced in \cite{Helou1985}.
\begin{equation}
    q_{\rm TIR}=\log{\bigg(\frac{L_{\rm TIR}}{3.75\times 10^{12}~{\rm W}}\bigg)}-\log{\bigg(\frac{L_{\rm 1.4~GHz}}{{\rm W~Hz^{-1}}}\bigg)}
\end{equation}
Some studies argued that this relation shows decrease with redshift increase \citep[e.g.,][]{2015A&A...573A..45M, 2017MNRAS.469.3468C, 2020MNRAS.491.5911O} which indicates the radio luminosity increases with redshift relative to IR luminosity.
We assumed redshift dependent total infrared to radio luminosity ratio as $q_{\rm TIR}(z)=(2.78\pm 0.02)\times (1+z)^{-0.14\pm 0.01}$ obtained by \cite{2017A&A...602A...4D} for VLA-COSMOS SFG sample.
For comparison, we applied constant $q_{\rm TIR}$ as $2.64\pm 0.02$ derived in \cite{2003ApJS..149..289B}.
This relation reflects the mechanisms of radio emission \citep{Yun2001, 2003ApJS..149..289B}. 
As for the conversion from IR luminosity to SFR, we adopt the relation based on \cite{KennicuttJr.1998}.
\begin{equation}
    {\rm SFRD}~[M_{\odot}\ {\rm yr}^{-1}\ {\rm Mpc^{-3}}]=\frac{\rho_{\rm IR}[L_{\odot}\ {\rm Mpc^{-3}}]}{5.8\times 10^9}
\end{equation}
where $\rho_{\rm IR}$ is IR luminosity density per unit comoving volume $[\rm Mpc^3]$ at redshift $z$ determined by integrating $\int^{L_{\rm max}}_{L_{\rm min}} L \phi(L,z)d\log L$. 
We set $(L_{\rm min},L_{\rm max})=(0,\infty)$ in the integration. While the integration range covers unrealistic values, the bulk of contribution to luminosity density comes from characteristic luminosity. 
Note that this assumption requires considerable extrapolation in LF at faint end especially in the high redshift as the data is limited to the bright end.
\begin{equation}
    {\rm SFRD}~[M_{\odot}\ {\rm yr}^{-1}\ {\rm Mpc^{-3}}]=f_{\rm IMF}\times 10^{q_{\rm TIR}(z)-24}{\bigg(\frac{3.0}{1.4}\bigg)}^{-\beta_{\rm SFG}}\frac{L_{3~{\rm GHz}}}{\rm [W~GHz^{-1}]}
\end{equation}
where $\beta_{\rm SFG}$ is the spectral index for SFG radio spectrum. 
The value of the coefficient $f_{\rm IMF}$ depends on IMF. In this study, we assumed Chabrier IMF \cite{2003PASP..115..763C}, thus $f_{\rm IMF}=1$.

The resultant cosmic SFRD is displayed in Fig.~\ref{fig:sfrd} with blue solid line along with the observational results from other wavelength. 
The shaded area corresponds to $95\%$ confidence interval of the LF resultant parameters. The SFRDs obtained by \citep{Smolcic2009, Karim2011, 2017A&A...602A...5N} are derived through radio surveys. 
\cite{Smolcic2009} SFRD, red diamonds, is based on LF in $1.4~{\rm GHz}$ estimated from 340 radio selected SFG up to $z=1.3$ detected in the VLA-COSMOS $1.4~{\rm GHz}$ survey \citep{Schinnerer2006}. 
\cite{Karim2011} constructed SFRD evolution model derived from stacked $1.4~{\rm GHz}$ continuum for $3.6~{\rm \mu m}$ selected SFG in the COSMOS field at $0.2<z<3$ which is plotted as green squares. 
While our result is baed on the same sample as \cite{Novak2018} (blue circles), their LF is based on $1/V_{\rm max}$ method. 
In order to compare our result with SFRD derived through LF in various wave band, we plot result obtained by \cite{Burgarella2013, Bouwens2015, 2017MNRAS.471.4155K}, yellow triangle, green stars and black dot-dashed line, respectively, and analytical form of the cosmic SFRD history obtained by \cite{Madau2014} with red dashed line. 
We also compare the result from UV band. \cite{Bouwens2015} derived rest-frame dust corrected UV LF obtained by over $10^4$ Lyman break galaxies at $4<z<10$ in the Cosmic Assembly Near-infrared Deep Extragalactic Legacy Survey (CANDELS) field identified by {\it HST} observation and \cite{Burgarella2013} combination of FIR ({\it Herschel}/PEP and {\it Herschel}/HerMES) and FUV LFs (VIMOS-VLT Deep Survey). 
\cite{2017MNRAS.471.4155K} estimated SFRD with SCUBA-2 and ALMA imaging data in sub-mm/mm up to $z\sim 5$ and adding the SFRD contribution derived by UV luminosity. 
\cite{Madau2014} compiled IR and UV data from multiple literature and derived parametric SFRD evolution function.

The resultant monotonically growing SFRD up to $z\sim 2.5$ agrees well with the literature. 
Especially, the SFRD by \cite{Novak2018} has a good agreement with our result for all redshift range considered and confirmed that the LF from independent method gives consistent evolution. 
The difference between our SFRD and \cite{Karim2011} becomes larger at $z>2$. The difference comes from the fact that they applied constant $q_{\rm TIR}$ parameter derived by \cite{2003ApJS..149..289B} which makes the IR luminosity larger at high redshift.
The brown dotted line indicates SFRD obtained assuming constant $q_{\rm TIR}$. 
While the result is has good agreement with \cite{Karim2011} and \cite{Burgarella2013} up to $z\sim 3$, the values at higher redshift is too high comparing to the literature. 
Adding to this, the model takes peak value at $z\sim 4$ which is too large.
Based on this result, redshift dependent $q_{\rm TIR}$ is gives SFRD which is consistent with the result from observations in other wavelength. 
We also plot SFRD whose integral base is assumed to be $L_{\rm lim}(z,S_{\rm lim})$ instead of $L_{\rm lim}=0$ with light-blue dotted line as a lower limit of prediction. Note that there is still uncertainty in the faint end slope to be constrained through future observation.

\section{The effect of gravitational lensing on galaxy number counts}
\label{sec:GL}
\begin{figure}
\begin{center}
\includegraphics[width=\columnwidth]{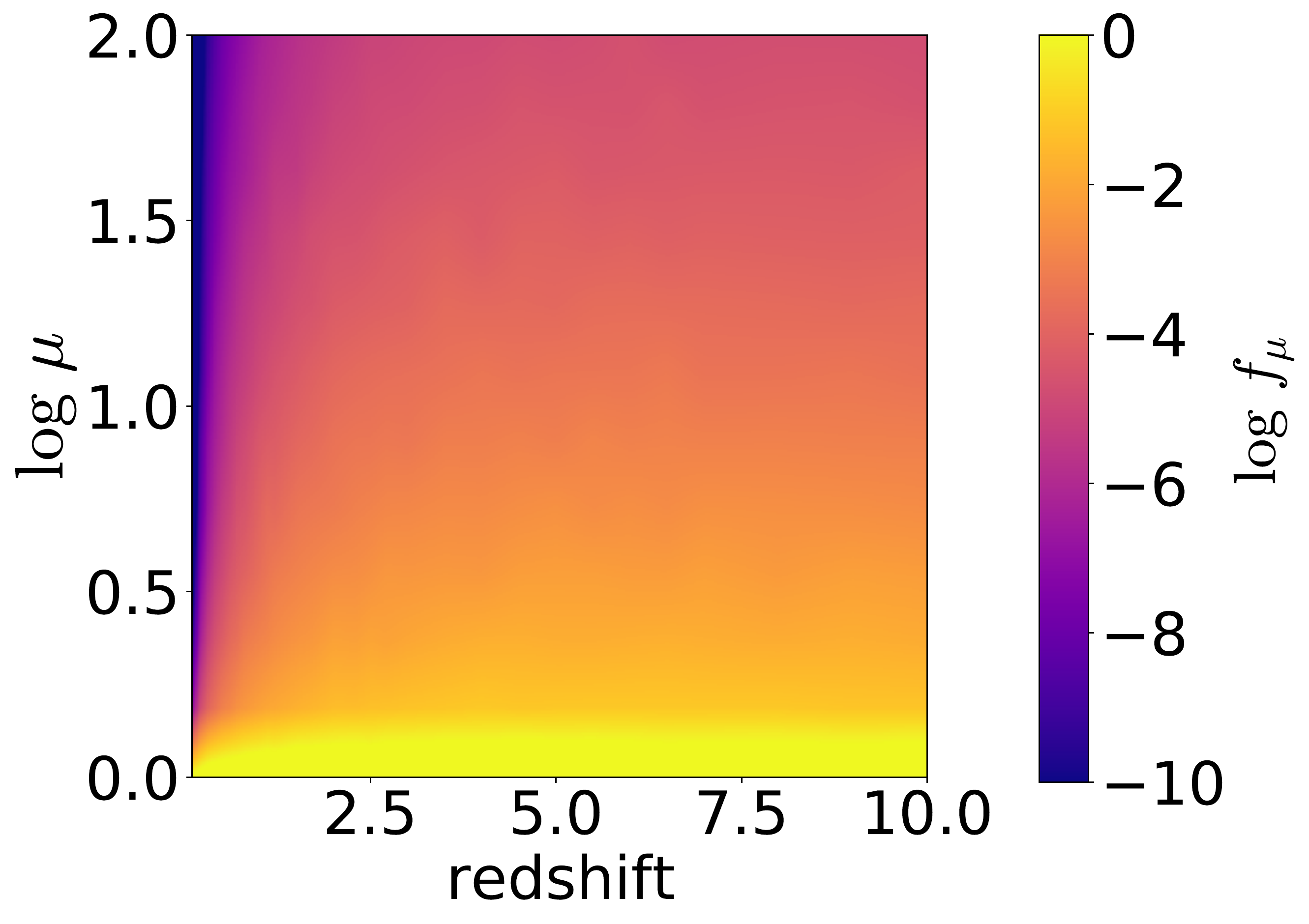}
\end{center}
\caption{The map of lensing probability $f_{\mu}$ estimated from Eq.~\ref{eq:fmu} as a function of source redshift and logarithmic magnification factor.}
\label{fig:map}
\end{figure}

Having constructed LF evolution model, we estimated the effect of gravitational lensing on galaxy number counts. The magnification $\mu$ due to gravitational lensing is given as the Jacobian between image coordinate and lens coordinate. Magnification profile $\mu(\theta)$ is calculated as below.
\begin{equation}
\label{eq:mu}
    \mu(\theta)=\frac{1}{(1-\kappa)^2-|\gamma|^2}
\end{equation}
where $\kappa$ is convergence, $\gamma$ is shear and $\theta$ denotes the separate angle from the center of the dark matter halo. 
These values depend on dark matter halo properties such as the distance to lens object $z_l$, mass $M$ and ellipticity $\epsilon$, and the distance to source object. 
We therefore adopt the parametric form of function from \cite{Sheth1999} for galaxy halo mass function and the NFW profile \citep{Navarro1997} as the radial density profile of each dark matter halo. We adopted the concentration parameter $c_{\rm vir}\equiv R_{\rm vir}/r_s$ in the NFW profile is given as a function of virial mass $M_{\rm vir}$ and redshift that is given by \cite{Bullock2001}. 
We also applied the analytical form for $\kappa$ and $\gamma$ obtained by  \cite{Takada2003a} and \cite{Takada2003b}.
Now we considered the ellipticity of DM halos $\epsilon$. It is well known that the ellipticity of the lens object modifies magnification profile. The ellipticity is defined as the ratio of projected semi-major $a$ and semi-minor axis $b$ as $\epsilon = 1-b/a$. 
In this elliptical coordinate, convergence and shear is modified as below. 
\begin{equation}
    \kappa_{\epsilon}(x)=\kappa(x_{\epsilon})+\cos{2\phi_{\epsilon}}\gamma(x_{\epsilon})
\end{equation}
\begin{equation}
    \gamma^2_{\epsilon}(x)=\gamma^2(x_{\epsilon})
    +2\epsilon \cos{2\phi_{\epsilon}}\gamma(x_{\epsilon})\kappa(x_{\epsilon})
    +\epsilon^2(\kappa^2(x_{\epsilon})-\cos^2{2\phi_{\epsilon}}\gamma^2(x_{\epsilon}))
\end{equation}
The elliptical coordinates are given as $x=\sqrt{x^2_{1\epsilon}+x^2_{2\epsilon}}$, $x_{1\epsilon}=\sqrt{a_{1\epsilon}}x_1$, $x_{2\epsilon}=\sqrt{a_{2\epsilon}}x_2$ and $\phi_{\epsilon}=\arctan{(x_{2\epsilon}/x_{1\epsilon})}$ which are defined in \cite{2002A&A...390..821G}, where $a_{1\epsilon}$ and $a_{2\epsilon}$ are the parameters defining the ellipticity in the lens coordinate. 
In this study, we applied $a_{1\epsilon}=1-\epsilon$ and $a_{2\epsilon}+\epsilon$. From this, we modify convergence and shear in eq.~(\ref{eq:mu}) as $\kappa \to \kappa_{\epsilon}$ and $\gamma \to \gamma_{\epsilon}$. 
We applied empirical value of ellipticity as $\langle \epsilon_{\rm DM} \rangle = 0.482\pm0.028$ estimated from {\it HST} cluster sample which is obtained by \cite{2020MNRAS.496.2591O}. 
In this study, we assumed triaxiality of dark matter halos does not depend on their mass for simplicity although they have mentioned that $\epsilon_{\rm DM}$ has lower value for lower $M_{\rm DM}$ based on simulation in \cite{2018MNRAS.478.1141O}.
In this study, we estimated effect of gravitational lensing on the statistical values of galaxy. 
We adopted the method from \cite{Lapi2012} and \cite{Lima2009}. 
The lensing probability $f(z_s)$ is computed as,
\begin{eqnarray}
\label{eq:fmu}
    f_{\mu}(z_s)&=&\int^{z_s}_0 dz_l 
    \int dM_{\rm H}  \\  
    & \times& \int d\epsilon P(\epsilon) \frac{dN}{dM_{\rm H}}\frac{d^2V}{dz_l d\Omega}\Delta \Omega(>\mu,M_{\rm H},z_s,z_l,\epsilon)     \notag
\end{eqnarray}
where $d^2V/dzd\Omega$ is the comoving volume element per unit solid angle and $dN/dM_{\rm H}$ is the galaxy halo mass function, $\epsilon$ is the ellipticity of dark matter halos and $P(\epsilon)$ is the probability distribution of the ellipticity.
We assumed that $P(\epsilon)$ to be $\delta\mbox{-}$function (i.e., $P(\epsilon)=\delta(\epsilon - \langle \epsilon_{\rm DM} \rangle)$) that only takes value at mean value obtained from observation. 
$\Delta \Omega$ in the third integration denotes the lensing cross section of a lens halo with $M_{\rm H}$ at $z_{l}$ for a source galaxy at $z_s$ whose flux is magnified by larger than $\mu$ times the original. 
The lensing cross section is derived by integrating the inverse of magnification with image plane coordinate considering the fact that the magnification is the Jacobian of source and image plane,
\begin{equation}
    \Delta \Omega\left(>\mu_{\min }\right)=\int_{\mu>\mu_{\min }} d \beta^{2}=\int_{\mu(\theta)>\mu_{\min }} \frac{d \theta^{2}}{\mu(\theta)}
\end{equation}
where $\beta$ and $\theta$ denotes the coordinates of the source plane and image plane respectively. 
Our logarithmic $f(z_s)$ is displayed in Fig.~\ref{fig:map} as a 2D map on source redshift $z_s$ and magnification $\mu$. Lensing probability decrease with $z_s$ as the possibility of having dark matter halo between the observer and source always decrease as the light path goes shorter. Especially, $f_{\mu}$ is significantly small  The $f_{\mu}$ at $\mu=2$ is approximately $1\times 10^{-2}$ and monotonically decrease with $\mu$ for above $z_s=0.2$. 

\begin{figure}
\begin{center}
\includegraphics[width=\columnwidth]{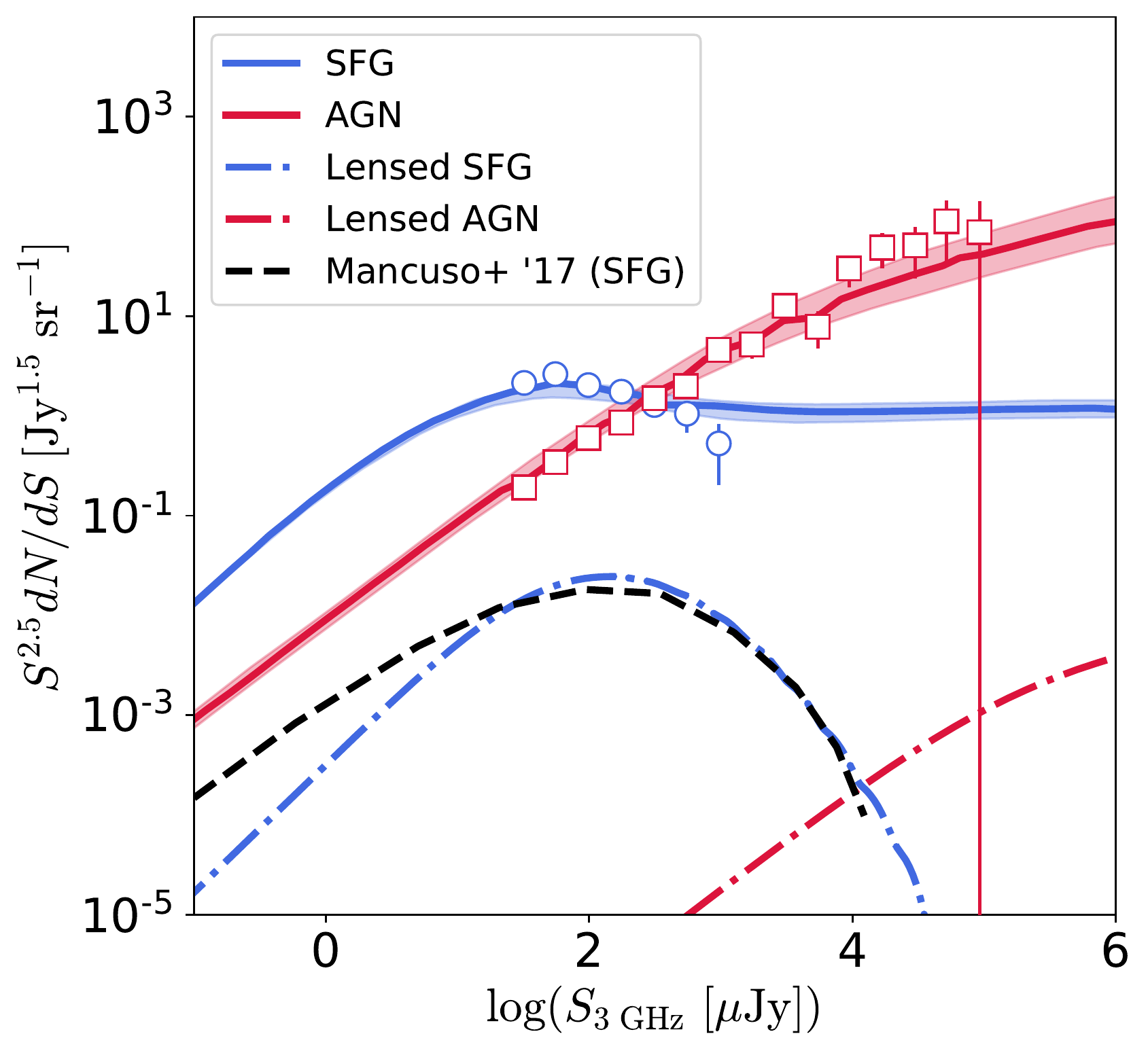}
\caption{The effect of gravitational lensing on the Euclid normalized galaxy differential number counts. 
The blue solid lines and circles, and red lines and squares show the original unlensed differential number counts for SFG and AGN, and the dot-dashed lines show the lensed differential number counts respectively.The black dashed line indicated lensed SFG derived by \citet{2017ApJ...842...95M} converted from $1.4~{\rm GHz}$ result.}
\label{fig:lense}
\end{center}
\end{figure}

Finally, the lensed differential number counts would be,
\begin{equation}
    \frac{dN}{dS}(S)=\int dz_s \frac{1}{\langle \mu \rangle} \int d\mu f_{\mu}(z_s) \frac{dn}{dS}{\bigg(\frac{S}{\mu},z_s\bigg)}
\end{equation}
where $dn/dS$ is differential source counts at $z=z_s$ whose flux is magnified by $\mu$ and $\langle \mu \rangle$ indicates average $\mu$ weighted by lensing probability.
Fig.~\ref{fig:lense} shows the result of analysis of the effect of gravitational lensing on the galaxy differential number counts. This result shows number counts is biased by gravitational lensing approximately $1\%$ at most in radio band with a peak around $S_{\rm 3GHz}\sim 100~{\rm \mu Jy}$. For SFGs, the effect would be diminished and becomes negligible towards bright end. This reflects the fact that the number counts of SFGs in bright end is dominated by contribution from nearby universe and lensing probability goes small at low $z_s$. We compare our result with lensed SFG differential number counts model obtained by \citet{2017ApJ...842...95M} which is constructed based on redshift dependent SFR function. While the peak flux agrees well with their result, our result shows subtler effect on lensed SFG counts especially towards faint population. On the other hand, the behaviour of lensed AGN differential number counts keeps growing trend because the population from high redshift is still considerable as discussed in Sec.~\ref{Sec:nc} for AGN. Nevertheless, the effect is smaller than that of SFG.

\subsection{Fisher analysis}
\begin{table*}
    \centering
    \begin{tabular}{c||ccc} 
               & Area $\mathcal{A}_{\rm obs}~[{\rm deg}^2]$ & Sensitivity $[{\rm \mu Jy/b}]$ & Frequency $\nu_{\rm obs}\ [{\rm GHz}]$ \\ \hline
    All-sky    & $3.1\times 10^4$     & 20.0                           & 20 \\ \hline
    Wide       & $1.0\times 10^3$     & 1.0                            & 1  \\ \hline
    Deep       & 10-30                & 0.2                            & 1  \\ \hline
    Ultra-deep & 1                    & 0.05                           & 1  \\ \hline
    \end{tabular}
    \caption{Outline of the four surveys planned in SKA1-MID.}
    \label{tab:SKA1}
\end{table*}
\begin{figure*}
\begin{center}
\includegraphics[width=\textwidth]{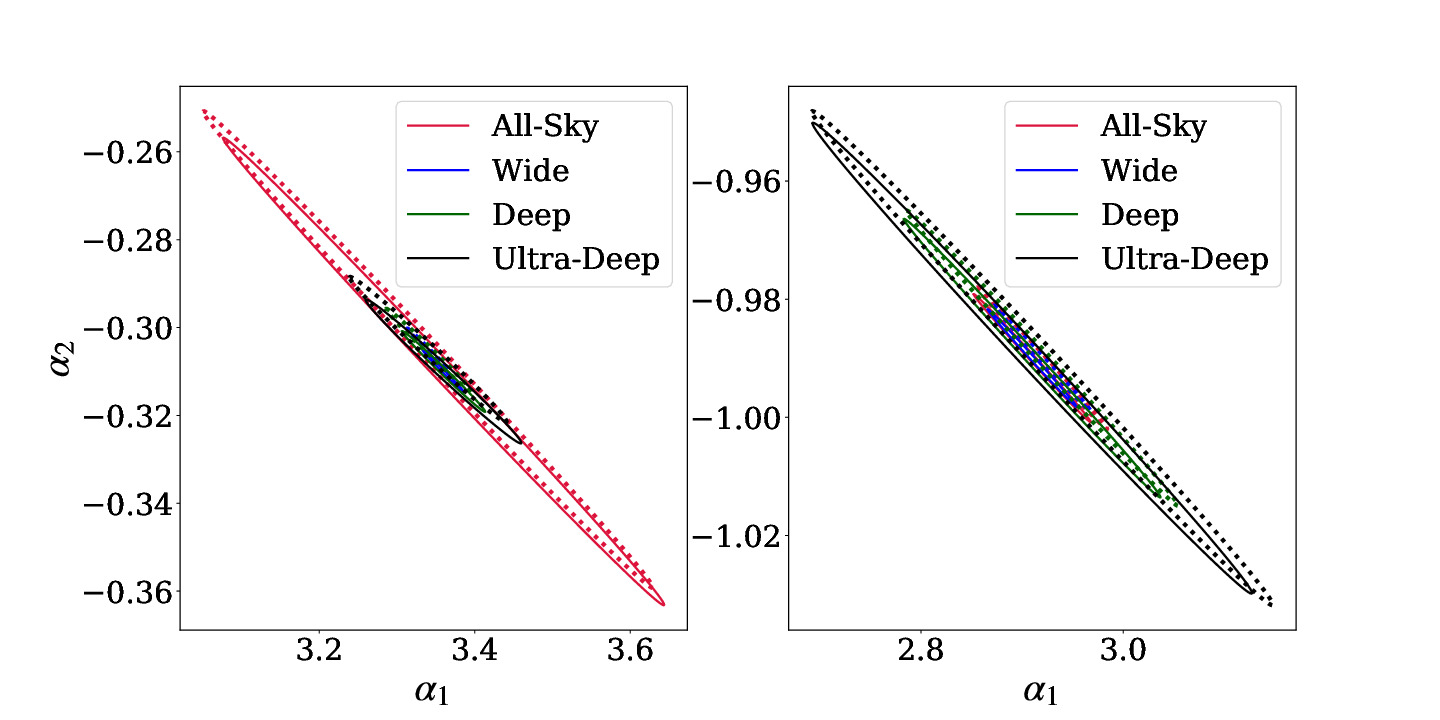}
\includegraphics[width=\textwidth]{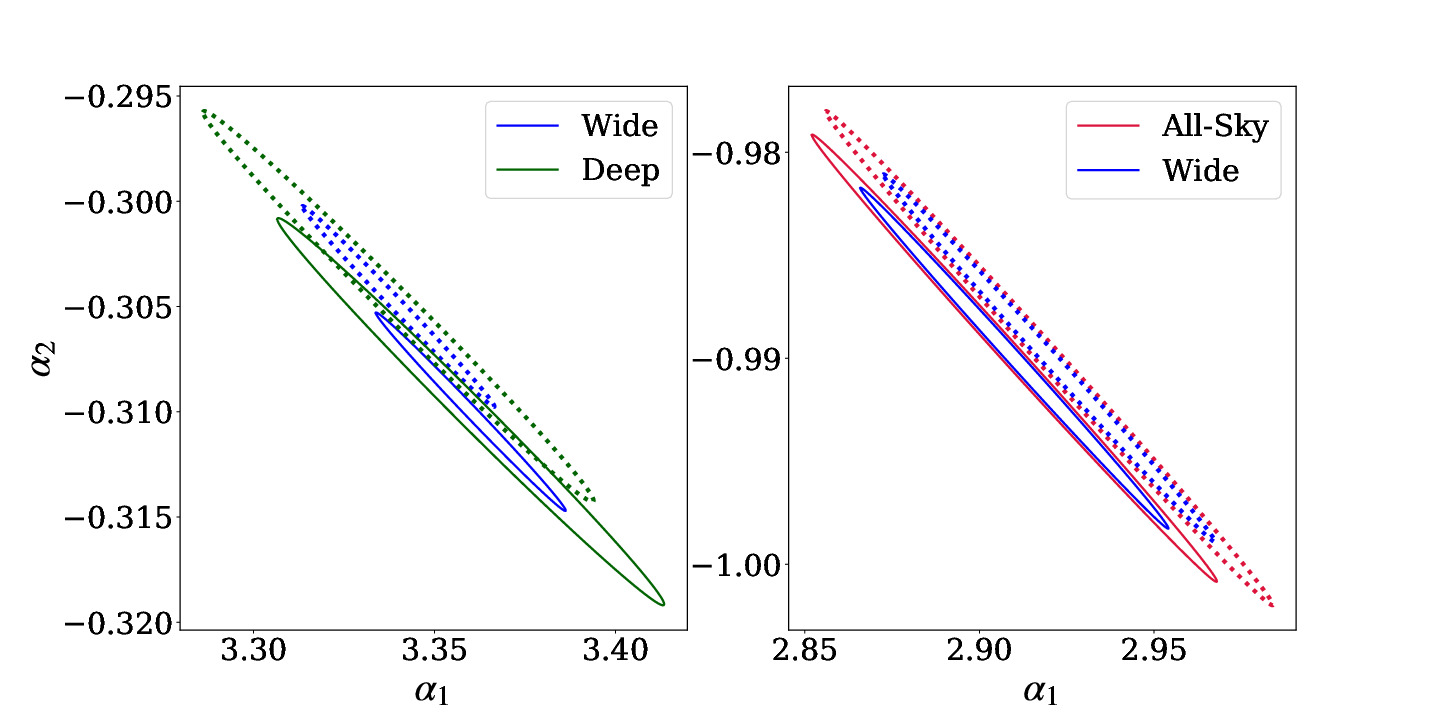}
\caption{Constrains of PLE evolution parameter $\alpha_1$ and $\alpha_2$ with Fisher analysis for SFG (left) and for AGN (right). Top figures are for all survey configurations and bottom figures are zooming up to the central region of the top figures.}
\label{fig:Fisher}
\end{center}
\end{figure*}
In order to investigate effect of the bias in estimation LF parameters from future surveys due to gravity lensing magnification, we performed Fisher analysis.  The expected lensed number count in $i,\ j$-th bin,
$N_{i,j}$, is given by
\begin{equation}
N_{ij} = 4 \pi \frac{\mathcal{A}_{\rm obs}}{\mathcal{A}_{\rm all}}
\int _{z_{i,\rm min}}^{z_{i,\rm max}} dz \int_{L_{{\rm min}},j}^{L_{{\rm max}},j} dL~
\frac{R^2(z)}{H(z)}
\frac{dn}{dL}{\left(\frac{S}{\mu}\right)},
\end{equation} 
where $L_{\rm min}$ is the luminosity corresponding to the sensitivity of four surveys with SKA. 
Here, the observed flux is magnified as $S_{\rm obs}=\mu S$. 
where $R(z)$ is the comoving distance to the redshift~$z$, ${z_{i,\rm min}} $ and $ {z_{i,\rm max}}$ are the minimum and maximum redshifts in the $i$-th redshift bin, $\mathcal{A}_{\rm obs}$ is the sky area of the observation.

In this analysis, we considered configurations of four large surveys which is planned in SKA-1 MID \cite{Prandoni2014}. 
The sensitivities, observation areas and observation frequencies for these survey designs are displayed in Tab.~\ref{tab:SKA1}. We assumed lensing fiducial PLE parameters that reproduces lensed number counts as $(\alpha_{1}^{\rm SFG,lensed},\alpha_{2}^{\rm SFG,lensed})=(3.36,-0.305)$ for lensed SFG number counts and $(\alpha_{1}^{\rm AGN,lensed},\alpha_{2}^{\rm AGN,lensed})=(2.92,-0.99)$ for lensed AGN, respectively.

The result of our parameter constraints with SKA surveys are displayed in Fig.~\ref{fig:Fisher} and predicted errors are summarised in Tab.~\ref{tab:delta_list}. The elliptical corresponds to $2\sigma$ errors. In the top two figures, we display constraints for all survey design. 
For SFG, all-sky survey gives the worst constraints. 
This is because, as indicated in Fig.~\ref{fig:diff}, high redshift SFG population dominates faint radio sky. 
Additionally, the difference in observing frequency causes the galaxy signal fainter due to negative spectral index. 
Thus, all-sky survey gives worse constraint than the VLA-COSMOS survey. 
The bottom-left panel in Fig.~\ref{fig:Fisher} shows constraints from wide and deep survey.
These two designs give stronger constrains on LF parameters than what is obtained in this study. Especially, the parameters for lensing fiducial model is distinguishable in the wide survey with $2\sigma$ error.
In the case of AGN parameter constraints, ultra-deep survey gives the worst constrains. As the AGN density parameter $\phi_*$ is about two orders of magnitude smaller than that of SFG, the small observation area critically decrease the number of AGN detection. As a result, the range of error elliptical for ultra-deep survey is comparable to the constraints from VLA observation. On the other hand, the lens fiducial parameter is distinguishable when operating all-sky and wide survey, and the parameters are constraint tighter by factor of two with these two surveys comparing with that of VLA-COSMOS survey. As summarized in Tab.\ref{tab:delta_list}, SKA I MID surveys gives tighter constrains on PLE parameters in most cases and deep survey is critically demanded to constraint the evolution for SFGs.

\begin{table}
    \centering
    \begin{tabular}{|c|c|cccc|}\hline
    \multicolumn{2}{|c|}{}
        & All-sky & Wide & Deep & Ultra-deep \\ \hline
        SFG & $\delta \alpha_1$ & $3.2\times 10^{-1}$ & $1.7\times 10^{-2}$ & $4.7\times 10^{-2}$ & $9.8\times 10^{-2}$ \\
            & $\delta \alpha_2$ & $5.1\times 10^{-2}$ & $5.0\times 10^{-3}$ & $1.8\times 10^{-2}$ & $3.0\times 10^{-2}$ \\ \hline
        AGN & $\delta \alpha_1$ & $4.0\times 10^{-2}$ & $3.0\times 10^{-2}$ & $8.0\times 10^{-2}$ & $2.0\times 10^{-1}$ \\
            & $\delta \alpha_2$ & $1.1\times 10^{-2}$ & $8.5\times 10^{-3}$ & $2.2\times 10^{-2}$& $4.0\times 10^{-2}$ \\ \hline
    \end{tabular}
    \caption{A summary of parameter constraints with SKA I MID galaxy surveys obtained by Fisher analysis.}
    \label{tab:delta_list}
\end{table}

\section{Summary}
\label{sec:conclusion}
In this paper, we have estimated the 3GHz LFs up to $z\sim 5.5$ for sources brighter than $S_{\rm 3GHz}>11.0\ {\rm \mu Jy}$ with making use of the data from the VLA-COSMOS 3GHz Large project, one of the deepest radio survey. 
The LFs are estimated through two independent method, a non-parametric $C^-$ method and parametric MCMC method. 
The sample contains 5410 SFGs and 1908 AGNs classified based on panchromatic counterparts and SFR based radio excess.
The evolution of the LF is described by PLE model.
The resultant redshift dependency of the characteristic luminosity is, $L_{*,{\rm SFG}}(z)\propto (1+z)^{(3.36\pm0.05)-(0.31\pm0.02)z}$ for SFG and $L_{*,{\rm AGN}}(z)\propto (1+z)^{(2.91^{+0.18}_{-0.19})-(0.99\pm0.08)z}$ for AGN. The 0th and 1st order term of PLE function showed a strong negative correlation. 
As expected from previous works, SFG shows stronger positive evolution comparing to that of AGN.
Our resultant LFs agreed well with that derived by \cite{Novak2018} with MCMC to determine the parameters of total LF obtained via non-parametric $1/V_{\rm max}$ method for overall redshift (Fig.~\ref{fig:LF}). The Euclidean normalized number counts estimated from our PLE model had a good agreement with various observations including ultra-deep surveys below $S_{\rm lim}$ in the VLA-COSMOS 3~GHz large survey while we assumed fixed faint end slope for LF (Fig.~\ref{fig:diff}).

We discussed about cosmic star formation rated density history estimated from radio luminosity.
Our SFRD model showed monotonic increase towards $z\sim 3$ and then compared our model with other SFRD models constructed from various wavelength in Fig.~\ref{fig:sfrd}.
We examined SFRD evaluated from constant radio to IR luminosity fraction for all redshift and found that constant $q_{\rm TIR}$ gives higher SFRD at $z>3.5$. \cite{2021ApJ...914..126M} introduced luminosity dependent non-linear $q$ parameter in order to exclude the bias caused by flux limited sample. The FIR/radio correlation is a critical quantity to draw SFRD from radio luminosity. It remained as an issue to be discussed further.

In the end, we assessed the statistical effect of gravitational lensing due to the dark matter halos with the estimated redshift distribution of galaxies. 
In the Euclidean normalized number counts, the lensed SFG population shows a peak at $S_{\rm 3GHz}\sim100\ {\rm \mu Jy}$ with a contribution of about $1\%$ while lensed AGN counts found to be less effective (Fig.~\ref{fig:lense}).  
Additionally, we performed Fisher analysis to predict the capability of PLE parameters for SFG and AGN parameter constraint with SKA I MID surveys. The direction of main axes of the error elliptical agrees with that is obtained for VLA-COSMOS sample (Fig.~\ref{fig:Fisher}) and we find that the parameter will be constrain about 10 times tighter in the SKA era (Tab.\ref{tab:delta_list}).

\section*{Acknowledgements}
This work has been supported by the Japan Society for the Promotion of Science (JSPS) Grants-in-Aid for Scientific Research (19H05076 and 21H01128). 
This work has also been supported in part by the Collaboration Funding of the Institute of Statistical Mathematics ``New Development of the Studies on Galaxy Evolution with a Method of Data Science''. 
\section*{Data Availability}

The data appeared in this article will be shared on reasonable request to the corresponding author.



\bibliographystyle{mnras}
\bibliography{MyCollection} 








\bsp	
\label{lastpage}
\end{document}